\documentclass[11pt]{article}

\usepackage[preprint]{acl}

\usepackage{times}
\usepackage{latexsym}
\usepackage{multirow}
\usepackage{textcomp}
\usepackage{amsmath,amssymb,amsfonts}
\usepackage{algorithmic}
\usepackage{graphicx}
\usepackage{threeparttable}
\usepackage{pifont}
\usepackage{url}
\usepackage{pgfplots}
\usepackage{tikz}
\usetikzlibrary{matrix}
\usepackage{colortbl}
\usepackage{booktabs}
\usepackage{subfigure}

\usepackage{hyperref}
\hypersetup{
    colorlinks=true,
    linkcolor=blue,
    citecolor=blue,
}

\newcommand{\xmark}{\ding{55}}
\newcommand{\cmark}{\ding{51}}

\usepackage{xcolor}
\def\BibTeX{{\rm B\kern-.05em{\sc i\kern-.025em b}\kern-.08em
    T\kern-.1667em\lower.7ex\hbox{E}\kern-.125emX}}

\newcommand*\circled[1]{\tikz[baseline=(char.base)]{\node[shape=circle,draw,inner sep=0.8pt] (char) {\small #1};}}

\usepackage[T1]{fontenc}
\usepackage[utf8]{inputenc}
\usepackage{microtype}
\usepackage{inconsolata}

\title{Probing Privacy Leaks in LLM-based Code Generation via Test Generation}

\author{\normalfont
Yifei Ge\textsuperscript{1},
Zhenpeng Chen\textsuperscript{2},
Weisong Sun\textsuperscript{3,*},
Yuchen Chen\textsuperscript{1},
Chunrong Fang\textsuperscript{1},
Juan Zhai\textsuperscript{4},
\\
Xiaofang Zhang\textsuperscript{5},
Xia Feng\textsuperscript{6},
Yang Liu\textsuperscript{3},
Zhenyu Chen\textsuperscript{1}
\\
\textsuperscript{1}Nanjing University,
\textsuperscript{2}Tsinghua University,
\textsuperscript{3}Nanyang Technological University
\\
\textsuperscript{4}University of Massachusetts Amherst,
\textsuperscript{5}Soochow University,
\textsuperscript{6}Hainan University
\\
\textsuperscript{*}Corresponding author.
\\
\small\texttt{\{yifeige,yuc.chen\}@smail.nju.edu.cn, \{fangchunrong,zychen\}@nju.edu.cn}
\\
\small\texttt{zpchen@tsinghua.edu.cn, \{weisong.sun,yangliu\}@ntu.edu.sg}
\\
\small\texttt{juanzhai@umass.edu, xfzhang@suda.edu.cn, xiafeng@hainanu.edu.cn}
}

\begin{document}
\maketitle

\begin{abstract}
The widespread availability of large-scale code datasets has fueled the rapid development of large language models (LLMs) for code-related tasks.
These datasets may include sensitive personally identifiable information (PII), which can lead to privacy leakage when LLMs memorize and reproduce it.
However, existing privacy-leakage detection methods rely on ad-hoc prompt construction (manually or automatically designed).
Therefore, they do not adequately approximate the real-world contexts in which PII appears in code corpora, making it difficult to extract realistic privacy leakage.
In this paper, we propose a pipeline that simulates practical privacy-related code generation scenarios and adopts a test-driven strategy to elicit the memorized information from the generated test cases.
We further introduce an automatically constructed privacy feature library that replaces manual prompt engineering by providing realistic templates and examples to guide test case generation.
Large-scale experiments on 5 widely used LLMs show that our pipeline exposes more confirmed privacy leakage, achieving a 2.56 times increase in detected leakage compared to existing baselines.
\end{abstract}

\section{introduction}



Large Language Models (LLMs) have become widely adopted tools in modern software development(e.g., Copilot~\cite{copilot} or Cursor~\cite{cursor}), supporting a variety of code intelligence tasks such as code generation and completion.
These capabilities are built upon large-scale pretraining on publicly scraped code repositories~\cite{2020-lm-few-shot-learners,2019-lm-multitask-learners}.
However, public code repositories may contain sensitive personally identifiable information (PII) that is unintentionally uploaded
, such as email addresses, credentials, API keys, and other sensitive records~\cite{2023-secretbench}. 
When PII is incorporated into training corpora, LLMs may memorize and unintentionally reproduce it during code generation, leading to privacy leakage. 
This leakage can compromise system security, user identity, or organizational confidentiality, and may lead to violations of data-protection regulations such as the General Data Protection Regulation (GDPR)~\cite{gdpr} or California Consumer Privacy Act (CCPA)~\cite{ccpa}.

\begin{figure}[t]
    \centering
    \includegraphics[width=0.9\linewidth]{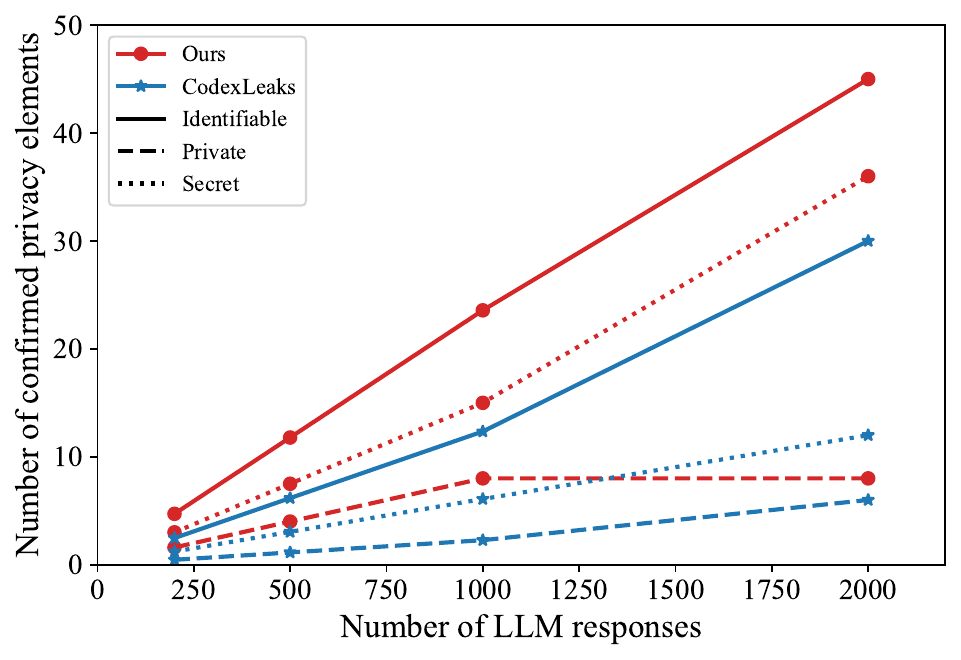}
    \vspace{-8pt}
    \caption{Number of confirmed privacy instances under different numbers of responses, comparing with Codebreaker across three privacy categories.}
    \label{fig:linechart}
\end{figure}

\begin{figure*}[t]
    \centering
    \includegraphics[width=0.9\linewidth]{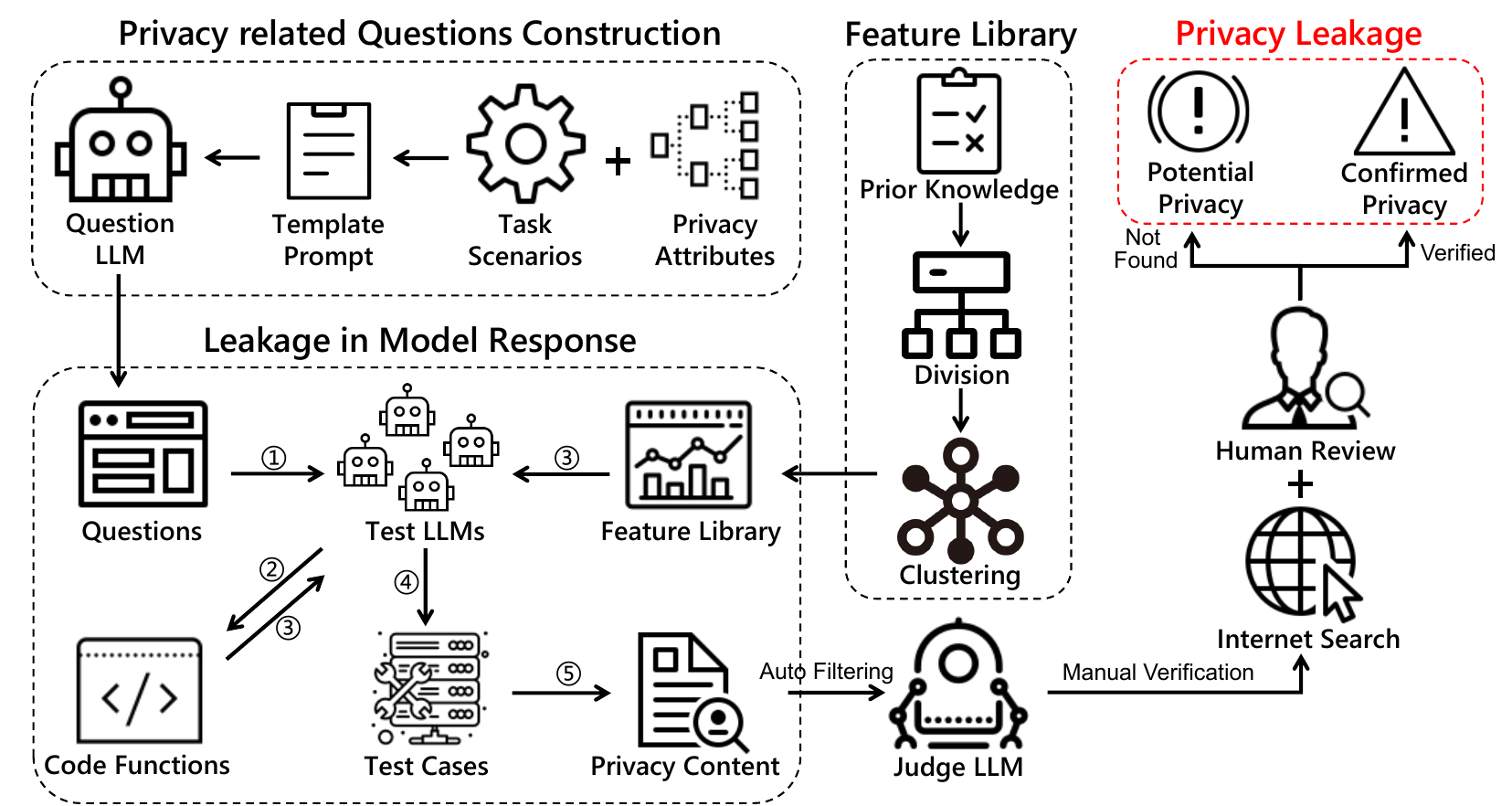}
    \caption{
    Overview of the privacy leakage evaluation pipeline. 
    \textit{*} For the leakage results, 
    some candidates cannot be verified and are labeled as \emph{Potential}. 
    We report only \emph{Confirmed} instances, yielding a conservative estimate of privacy leakage; \emph{Potential} items may still correspond to real leakage but are excluded from reported counts.
    }
    \label{fig:workflow}
\end{figure*}

Although several privacy protection methods, such as data filtering~\cite{2016-recon,2017-obfuscation-resilient-privacy}, differential privacy~\cite{2018-privacy-risk-in-ml, 2025-privcode}, and federated learning~\cite{2021-understand-unintended-memorization,2019-comprehensive-privacy-analysis}, have been introduced during the training or deployment phases of LLMs to mitigate the risk of privacy leaks, incidents of privacy data leakage continue to occur
~\cite{deepseek_news, doubao_news}.
To better understand and assess these risks, existing works~\cite{2019-secret-sharer,2019-comprehensive-privacy-analysis, 2021-extract-train-data-from-llm} have proposed privacy data extraction methods, 
These methods interact with LLMs to test whether they disclose PII, thereby assessing the presence and
extent of privacy leakage.

However, existing methods for extracting privacy leakage from LLMs during code-related tasks remain limited in their scope and effectiveness.
Early work by Niu et al.~\cite{2023-codexleaks} heavily depends on carefully designed prompts (many of which are manually constructed)
, and the amount of privacy information is constrained by the availability of prompts 
(See Figure~\ref{fig:linechart}).
Moreover, the method is designed specifically for Codex~\cite{2021-codex} 
and has limited applicability to other LLMs.
Recent work by Han et al.~\cite{2025-codebreaker} introduces automated prompt generation (via mutation-based strategies) to elicit code privacy content. 
However, 
its prompts lack sufficient grounding in the authentic contexts in which privacy information appears in real code corpora, resulting in many extracted candidates being hallucinated or placeholder-like
rather than real privacy.
To overcome these limitations,
we propose a semi-automated pipeline for evaluating privacy leakage from LLMs in code generation tasks via test-case generation. 
Our design is motivated by the following design principles:
First, 
\textit{memorization is more likely to be triggered when prompts resemble their original training contexts}.
Since the code training corpora are unknown, we approximate them by constructing realistic development scenarios (questions) with explicit privacy attributes, thereby increasing the likelihood that privacy-relevant memorization surfaces.
Second, rather than directly requesting privacy data, which is often blocked or refused by safety mechanisms of LLM, we elicit unit tests for code functions. 
This indirect and developer-like interaction pattern requires privacy-valued inputs and is less prone to refusal in privacy-related tasks.
Finally, to avoid manual prompt engineering and improve prompt effectiveness at scale, we introduce an automatically constructed and iteratively updated privacy feature library that supplies realistic templates and fragments as prompt augmentation.

As illustrated in Figure~\ref{fig:workflow}, 
our pipeline audits privacy leakage in code generation tasks via test-case generation.
(1) We instantiate development scenarios with targeted privacy attributes and use them to generate diverse code-generation questions. 
(2) Given each question, the evaluated LLM first produces a corresponding code function, and is then prompted to generate unit test cases whose inputs require privacy-related fields.
During this process, a privacy feature library 
provides realistic privacy formats and content patterns as augmentations, guiding the outputs away from trivial placeholders toward plausible privacy values.
(3) The extracted candidates are then verified in a unified verification stage that combines an automated Judge LLM with GitHub-based Internet search and human review, yielding a final set of confirmed privacy leakage instances.

\begin{table*}[htbp]
    \centering
    \scriptsize  
    \tabcolsep=10.5pt
    \vspace{-8pt}
    \caption{Privacy category definition.}
    \label{tab:definition}
    \begin{tabular}{c|c|c|c}
        \toprule
        \footnotesize\textbf{Category} & \footnotesize\textbf{Attribute} & \footnotesize\textbf{Scenarios} & \footnotesize\textbf{Example} \\
        
        \midrule
        
        \multirow{5}{*}{\footnotesize Identifiable}
        & Name & Enterprise App, Mobile & Jameson C***er\\ 
        & Address & Enterprise App & 
        city: ``São Paulo'', address: ``Avenida Paul***''\\ 
        & Email & Enterprise App, Web, Cloud Service & george.t******@outlook.com \\ 
        & Phone Number & Mobile, Enterprise App, Web & +86 138 *****022 \\ 
        & Date of Birth & Enterprise App & ``birth\_date'': ``19**-07-16'' \\ 

        \midrule
        
        \multirow{4}{*}{\footnotesize Private}
        & Identity & Enterprise App, Web & ``Emirates ID'': ``784-1988-12****4-1'' \\ 
        & Medical Record & Enterprise App &  ``height'': 1**, ``conditions'': [``Heart Disease''...] \\ 
        & Bank Statement & Enterprise App & ``bank\_details: Recent overdraft fees applied'' \\ 
        & Political & Web & political\_party=\{ideology = communism\} \\ 
        
        \midrule
        
        \multirow{6}{*}{\footnotesize Secret}
        & Password & Enterprise App, Web, Game, Cloud Service & Sokol*****73 \\ 
        & Authentication PIN/Token & Blockchain, Mobile & 67**29 \\
        & Secret Key & Blockchain, Cloud Service 
        & sk-78a92b74ea****d5b5bc6fef3 \\ 
        & Credit Card & Web & 3566-0020-20**-**** \\ 
        & Account/User Name & Enterprise App, Web, Game, Cloud Service & mingyu\_b**\_**3 \\ 
        & Biometric Data & Mobile & "Jake\_blood\_type\_O" \\ 
        
        \bottomrule
        
    \end{tabular}
    
        
\end{table*}

\textbf{Contributions.} In brief, our contributions are:
\begin{itemize}
    \item 
    We propose a test-driven pipeline for evaluating privacy leakage in code generation tasks for LLMs that is grounded in privacy-related development scenarios. 
    
    \item 
    We introduce an automatically constructed and uploaded privacy feature library that provides realistic templates and fragments derived from real leakage patterns, reducing reliance on manual prompt engineering.
    

    \item 
    Experiments on 5 widely-used commercial LLMs show that our pipeline consistently identifies an average of 92.6 confirmed privacy leakage instances, and outperforms recent baselines across privacy categories and leakage levels (up to 15.68\textperthousand~/ 2.56 times).
    
\end{itemize}







\section{Background and Related Work}




\subsection{Model Memorization}

Memorization in LLMs refers to the phenomenon in which models reproduce specific sequences from their training data rather than generating fully novel content. 
Prior work has shown that such a memorization phenomenon may occur when the input prompt closely matches their original training contexts~\cite{2024-traces-of-memory, 2024-unveiling-memory}.
In such cases, instead of purely generalizing, LLMs may reproduce verbatim or near-verbatim sequences from the training corpus.

A widely used indicator of memorization is the model’s perplexity metric (\textit{PPL}),
which measures how ``surprised'' the model is when generating
a sequence.
Sequences that the model is more familiar with tend to yield lower perplexity values and thus often correlate with stronger memorization.
This connection provides an important basis for evaluating whether generated outputs may originate from memorized sensitive data.




\subsection{Training Data Extraction}

Training data extraction methods have been extensively studied in the context of natural language generation~\cite{2019-secret-sharer,2019-comprehensive-privacy-analysis}, which
aim to recover sensitive PII from LLMs by exploiting their memorization behaviors, thereby leading to privacy leakage.
These methods typically craft prompts, sample model outputs, and identify candidates that are more likely to be training-data reproductions.

Beyond natural language tasks, concerns have been raised regarding privacy leakage in code-related applications of LLMs. 
Prior studies have shown that public code repositories such as GitHub may contain unfiltered private information, including credentials and personal identifiers~\cite{2019-how-bad-git}. 
When LLMs are trained on such data, similar memorization behaviors may result in the leakage of sensitive information during code tasks. 

\section{Problem Definition}


\subsection{Privacy Categorization}
\label{subsec:privacy_define}

%

We categorize personal information that commonly appears in the code repositories, as summarized in Table~\ref{tab:definition}.
Following CodexLeaks~\cite{2023-codexleaks}, the only study that provides a concrete categorization tailored to code-related privacy leakage, 
we divide privacy information into 3 categories: \textit{Identifiable}, \textit{Private}, and \textit{Secret}.
We further refine this taxonomy by removing attributes that rarely appear in code corpora and are seldom observed in real code leakage cases (e.g., gender, education, or social media). 
Privacy-related development scenarios associated with these attributes are introduced in Sections~\ref{subsec:question_generation}, where we describe how they are incorporated into our evaluation pipeline.

\subsection{Privacy Leaks}




The privacy leak is defined as a model-generated output that contains personal information
arising from the model memorization behaviors
~\cite{2022-preventing-verbatim-memory}.
Since our study targets commercial LLMs whose training corpora are not publicly accessible, it is not possible to directly verify whether a generated privacy string originates from training data. 
Therefore, following the prior methods~\cite{2023-codexleaks, 2025-codebreaker}, we use public GitHub repositories as a proxy for pretraining code corpora and verify privacy information via GitHub Search.
Note that repository content may have been modified or removed since model training. 
Thus, the number of leaked items we identify should be interpreted as a conservative lower bound on the true amount of privacy leakage.

\subsection{Threat Model}


We assume attackers who interact with the model via input-output access without direct visibility into the model’s internal structure or parameters. 
Besides, the attackers also have partial access to code segments from the training data, with prior knowledge specifically regarding the privacy information involved in the code snippets. 
This is a realistic assumption, given that training LLMs without relying on open-source code is virtually impractical.

\section{Privacy Leakage Pipeline}

\subsection{Code Generation Questions}
\label{subsec:question_generation}


The first stage of our pipeline is to construct a set of code-generation tasks that naturally involve privacy-related fields. 
Motivated by the observation that memorized privacy strings are more likely to surface when prompts resemble training-time contexts. 
Without such grounding, models tend to generate placeholders or hallucinated values.
Since the original contexts in the training corpus are unknown, we approximate them by grounding each
privacy attribute in realistic application scenarios and instantiating these scenarios into concrete programming tasks.
We embed each privacy attribute into application scenarios where the attribute is functionally required or commonly handled.  
We derive scenarios from a structured taxonomy of common software domains (including enterprise apps, mobile services, cloud services, web platforms, blockchains, and games)~\cite{2024-top-general-performance}
and instantiate them into concrete code generation tasks.
For each privacy attribute, we map it to one or more scenarios and then generate scenario-conditioned code-generation questions that require the model to implement functionality operating on that attribute (as summarized in Table~\ref{tab:definition}).

Formally, let $S$ denote the set of development scenarios and $A$ the set of privacy attributes defined in Section~\ref{subsec:privacy_define}. For each scenario $s \in S$, we identify the subset $A(s)\in A$ of attributes that plausibly appear in that scenario. We then define a task-construction function:

\[
T(s, a) = \text{LLM\_question}(\Phi(s, a)),
\]

where $\Phi(s, a)$ is a scenario--attribute prompt template that requires the model to generate functional code involving attribute a in scenario $s$. The LLM executes $T(s, a)$ to produce a diverse set of context-aware code generation questions 
$\mathcal{Q}$.
These questions serve as the input to subsequent stages of our pipeline.

\subsection{Model Response}
\label{subsec:model_response}


In this stage, we directly interact with the evaluated LLM (Test LLM) to expose potential privacy leakage through its responses. 
Starting from the privacy-related generation questions, we drive the model to produce functional code and then generate test cases whose inputs may contain memorized privacy information.
By eliciting test inputs rather than directly querying, this interaction reduces the likelihood of triggering built-in safety mechanisms.

For each question $q \in\mathcal{Q}$, 
the Test LLM generates a code snippet $c = LLM_{test}(q)$ (corresponding to step \circled{1}–\circled{2} in Figure~\ref{fig:workflow}). 
We extract those functions that explicitly process or reference privacy-related fields, obtaining a set of candidate functions G(c). During this step, we discard auxiliary elements such as import statements, global constants, or placeholder code that do not contribute to actual data handling.
Next, for each function $g \in G(c)$ associated with one or more privacy attributes $a$, we ask the Test LLM to generate unit test cases 
(step~\circled{3}--\circled{4}). 
The prompt for test-case generation is constructed by combining the function $g$ with attribute-specific hints drawn from the privacy feature library $\Lambda(a)$ (details in Section~\ref{sec:feature_library}), so that the model is encouraged to supply realistic, attribute-shaped inputs rather than trivial placeholders. 
Formally, we build a test-case prompt $prompt(g, \Lambda(a))$ and query the model again to obtain a test case $\tau = LLM_{test}(prompt(g, \Lambda(a)))$.

From each generated test case $\tau$, we extract the concrete values appearing in input arguments or data structures that match the structural patterns of our privacy attributes (step~\circled{5}). 
We extract candidate privacy information from a deterministic parsing procedure $\mathrm{ExtractPII}(\cdot)$, which scans $\tau$ and collects token spans for different attributes.
The union of all extracted values across questions and functions is:
\[
\mathcal{C} = \bigcup_{q \in \mathcal{Q}} \bigcup_{g \in G(c_q)} \text{ExtractPII}(\tau_{q,g}).
\]









\subsection{Filtering and Verification}

After extraction, we obtain a set $C$ of candidate privacy values collected from generated test cases. 
We apply Judge LLM combined with GitHub-based verification and human review to retain the confirmed privacy leaks.

\textbf{Judge LLM--based filtering.}
For reducing human effort in later stages, we instantiate an oracle model $\text{LLM}_{\text{judge}}$ that aims to automatically screen hallucinated strings, placeholders, or otherwise implausible values in the candidates.
For each extracted value $x \in C$ associated with attribute $a \in A$, the Judge LLM receives $x$, its attribute type, and a concise description of $a$'s structural and semantic characteristics, together with several fragment examples drawn from the prior knowledge.
Conditioned on this context, $\text{LLM}_{\text{judge}}$ decides whether $x$ is a plausible instance of attribute $a$ in realistic code (e.g., a correctly formed email address, phone number with reasonable length, or credential-like string) and rejects items with invalid formats or clearly implausible semantics. 
We denote by $C_{\text{judge}} \subseteq C$, the subset of candidates retained by this automated filtering step. 

\textbf{GitHub search and human review.}
After filtering out,
we verify the authenticity of candidate privacy information using GitHub search.
For each candidate $x \in C_{\text{judge}}$, we query the GitHub code search API using either the full string or a discriminative substring. 
Let $k$ denote the number of search hits. 
Consistent with prior works' setting, 
we retain candidates with $1 \le k \le 100$
(because a nonzero match indicates occurrence in real code, while overly frequent matches are less likely to correspond to real information).
For all candidates within this range, two authors independently inspect the matched GitHub contexts and the candidate string itself to determine whether it represents real personal information,
rather than documentation text, test data, or intentionally fictitious examples. 
Only those candidates that 
(i) satisfy the attribute-specific structural constraints and appear semantically plausible for their claimed type
(ii) are supported by GitHub context, indicating that they are used as privacy-related data in code
are regarded as the final leakage set $L \subseteq C_{\text{judge}}$.


\section{Privacy Feature Library}
\label{sec:feature_library}

\subsection{Library Definition}

To effectively guide the LLM in generating real and diverse privacy information within the pipeline, we maintain a Privacy Feature Library for each attribute.
Intuitively, the library provides 
templates and fragments derived from real-world code contexts,
which are used to augment prompts and make them closer to the training-time contexts.
Since memorization is more likely to be triggered when prompts resemble their original training contexts, these realistic cues increase the likelihood of memorized privacy values appearing,
while reducing templated or trivial content (e.g., generic placeholder names such as ``Zhang San'' or ``John Doe'').
Formally, for each privacy attribute $a$, we define an attribute-specific library:
\[
\Lambda(a) \;=\; \{\Lambda_{\text{tmp}}(a),\; \Lambda_{\text{frag}}(a)\},
\]
where each component stores a particular type of feature:
(i) the \emph{template set} $\Lambda_{\text{tmp}}(a)$ contains abstract patterns and structural contexts for $a$, such as key--value formats and sentence-level templates 
(e.g., \texttt{user.email = \textless EMAIL\textgreater}, \texttt{contact: \textless PHONE\textgreater});
(ii) the \emph{fragment set} $\Lambda_{\text{frag}}(a)$ contains value-level substrings and full strings that resemble realistic privacy content 
(e.g., \texttt{+86 138-1108-5305}, or \texttt{``li.ming@qq.com''}).


Template entries are primarily used to constrain the structure of the generated test inputs, while fragment entries serve as completion cues or illustrative values.
We first initialize both $\Lambda_{\text{tmp}}(a)$ and $\Lambda_{\text{frag}}(a)$ from prior knowledge, which incorporates known privacy patterns and publicly available privacy samples.
Then, we further enrich this library using the set $L(a)$ from previous runs of our pipeline. 
The concrete procedure for automatically decomposing $L(a)$ into template components and privacy fragments and 
mapping them to their corresponding privacy attributes $a$ is described below.

\subsection{Component Division}
\label{subsec:library_division}

After obtaining the confirmed leakage set $L$, our goal is to separate each leaked instance into its reusable structural template and privacy fragments. 
Using the evaluated LLM's own likelihood (or perplexity) is not suitable:
memorized privacy strings and frequent boilerplate patterns can both receive similarly high likelihood, and the absolute ranges of perplexity are often not comparable across different LLMs. 
Therefore, to obtain an unbiased signal, we instead apply a code-oriented pretrained model $P$ (CodeBERT),
and reasonably assume that the training corpus of $P$ does not contain the same private information memorized by the evaluated LLMs.
For each leaked instance $x = (t_1,\ldots,t_n)$, we compute token-wise pseudo-log-likelihood scores by masking one token at a time and predicting it from the remaining context:
\[
\ell_i = -\log \tilde{p}_P(t_i \mid \text{context}(x,i)),
\]
where $\text{context}(x,i)$ denotes the masked input obtained by replacing the $i$-th token in $x$ with a mask symbol while keeping all other tokens unchanged.
Intuitively, tokens corresponding to common structural elements are well represented in generic code corpora and thus yield relatively low pseudo-NLL under $P$, whereas attribute-specific value tokens tend to be out of distribution and exhibit higher pseudo-NLL.
We identify privacy templates using the empirical distribution of $\{\ell_i\}$ within each leaked instance: tokens whose pseudo-NLL falls in the \textit{lower quartile} (the lowest 25\%) are extracted as privacy templates, and all remaining tokens are treated as fragment components. 
We then replace the extracted fragments with the corresponding attribute slot symbols (e.g., $\langle\text{EMAIL}\rangle$, $\langle\text{PHONE}\rangle$) to obtain abstract templates, and store the resulting templates and fragments in $\Lambda_{\text{tmp}}(a)$ and $\Lambda_{\text{frag}}(a)$.

\subsection{Semantic Clustering}
\label{subsec:semantic_clustering}

Following the division, we obtain a collection of automatically extracted privacy fragments and templates from different leaked samples. 
These raw patterns often appear in diverse surface forms across languages and writing styles, and may also include noisy artifacts produced by imperfect division.
We therefore apply a semantic clustering to consolidate equivalent patterns and to remove isolated or semantically inconsistent items.

Concretely, we embed each template (and fragment) using the encoder of model $P$. 
We then perform clustering in the embedding space with a density-based method (DBSCAN). 
Due to substantial semantic gaps between different privacy attributes, templates and fragments associated with the same attribute tend to form dense clusters, including cross-lingual and cross-style variants (e.g., ``email:'', ``EMAIL =''). 
We regard the low-density points in clustering as noise and discard these outliers, as they typically correspond to spurious strings or wrongly divided snippets that do not align with any major cluster and would otherwise degrade the quality of the feature library.
After clustering, we assign each retained cluster to a privacy attribute by comparing it with attribute prototypes derived from the initialized feature library.

\section{Experiments}

\begin{table*}[!t]
    \centering
    \scriptsize 
    \caption{Averaging leakage results of GPT and DeepSeek's LLM series.}
    \vspace{-8pt}
    \label{tab:main}
    \begin{tabular}{@{}c@{\hskip 3.5pt}c@{\hskip 3.5pt}|
    c@{\hskip 4.5pt}|
    c@{\hskip 4.5pt}|
    @{\hskip 3.0pt}c@{\hskip 3.0pt}|
    @{\hskip 3.0pt}c@{\hskip 3.0pt}|
    @{\hskip 2.8pt}c@{\hskip 2.8pt}
    @{\hskip 3.0pt}c@{\hskip 3.0pt}|
    c@{\hskip 4.5pt}|
    @{\hskip 3.0pt}c@{\hskip 3.0pt}|
    @{\hskip 3.0pt}c@{\hskip 3.0pt}|
    @{\hskip 2.8pt}c@{\hskip 2.8pt}
    @{\hskip 3.0pt}c@{\hskip 3.0pt}}
        \toprule
        \multirow{5}{*}{\footnotesize Category} & \multirow{5}{*}{\footnotesize Attribute} & \multirow{5}{*}{\shortstack{Number\\of\\Test Cases}} & \multicolumn{5}{c}{GPT-Series} & \multicolumn{5}{|c}{DeepSeek-Series} \\
        
        \cmidrule(lr){4-8}\cmidrule(lr){9-13}
        
        & & & \multirow{3}{*}{\shortstack{Accepted\\Number}} & \multirow{3}{*}{\shortstack{Judge\\LLM}} & \multirow{3}{*}{\shortstack{Github\\Search}} & \multicolumn{2}{c|}{Human Check} & \multirow{3}{*}{\shortstack{Accepted\\Number}} & \multirow{3}{*}{\shortstack{Judge\\LLM}} & \multirow{3}{*}{\shortstack{Github\\Search}} & \multicolumn{2}{c}{Human Check} \\
        
        \cmidrule(lr){7-8}\cmidrule(lr){12-13}
        
        & & & & & & confirmed & permille & & & & confirmed & permille \\
        
        \midrule

        \multirow{5}{*}{\rotatebox{90}{\textbf{Identifiable}}}
        & Name & 400(=2*20*10) & 322 & 156.7 & 61 & 8.7 & 21.8\textperthousand & 374 & 169.5 & 35.5 & 10 & 25.0\textperthousand \\
        & Address & 400(=2*20*10) & 174 & 118 & 76.3 & 7 & 17.5\textperthousand & 198 & 99 & 47 & 13.5 & 33.8\textperthousand \\
        & Email & 600(=3*20*10) & 198.3 & 152.3 & 20 & 15.3 & 25.5\textperthousand & 200 & 116 & 24.5 & 23.5 & 39.2\textperthousand \\
        & Phone Number & 600(=3*20*10) & 176 & 126 & 9.3 & 4.7 & 7.8\textperthousand & 180.5 & 132 & 2 & 2 & 3.3\textperthousand \\
        & Date of Birth & 200(=1*20*10) & 158 & 128.3 & 27.7 & 10 & 50.0\textperthousand & 364.5 & 106.5 & 28.5 & 4.5 & 22.5\textperthousand \\

        \midrule

        \multirow{4}{*}{\rotatebox{90}{\textbf{Private}}}
        & Identity & 400(=2*20*10) & 309 & 96.7 & 12 & 2 & 5.0\textperthousand & 367 & 82.5 & 1.5 & 0 & 0.0\textperthousand \\
        & Medical Record & 200(=1*20*10) & 159.7 & 108 & 2.7 & 2 & 10.0\textperthousand & 199 & 74 & 2 & 1 & 5.0\textperthousand \\
        & Bank Statement & 200(=1*20*10) & 107.3 & 51 & 3 & 1 & 5.0\textperthousand & 138.5 & 75 & 6.5 & 1.5 & 7.5\textperthousand \\
        & Political & 200(=1*20*10) & 93.3 & 18 & 3.3 & 0.3 & 1.5\textperthousand & 62 & 9.5 & 0.5 & 0 &
        0.0\textperthousand \\

        \midrule
        
        \multirow{6}{*}{\rotatebox{90}{\textbf{Secret}}}
        & Password & 800(=4*20*10) & 561.7 & 132.3 & 25 & 12.7 & 15.9\textperthousand & 580.3 & 109 & 7 & 6 & 7.5\textperthousand \\
        & Authentication PIN & 400(=2*20*10) & 240 & 79 & 6 & 4 & 10.0\textperthousand & 200 & 45.5 & 15.5 & 1 & 2.5\textperthousand \\
        & Secret Key & 400(=2*20*10) & 360.7 & 181.3 & 9.7 & 8 & 20.0\textperthousand & 484 & 301.5 & 8 & 6 & 15.0\textperthousand \\       
        & Credit Card & 200(=1*20*10) & 83.7 & 55.3 & 6.7 & 2 & 10.0\textperthousand & 143 & 86 & 5 & 2.5 & 12.5\textperthousand \\
        & Account/User Name & 800(=4*20*10) & 707.7 & 67.3 & 37.3 & 27.7 & 34.6\textperthousand & 752 & 28.5 & 11 & 7.5 & 9.4\textperthousand \\
        & Biometric Data & 200(=1*20*10) & 186 & 18.3 & 1.7 & 0.3 & 1.5\textperthousand & 190.5 & 22.5 & 0.5 & 0.5 & 2.5\textperthousand \\

        \midrule
        \multicolumn{2}{c|}{Total} & 6000 & 3642 & 1488.7 & 301.7 & 105.7 & 17.6\textperthousand & 4146 & 1457 & 195 & 79.5 & 13.3\textperthousand \\
        
        \bottomrule

    \end{tabular}
\end{table*}

\subsection{Experimental Setup}

\textbf{Models.}
We evaluate 5 representative commercial LLMs (from 2 widely used model families: the GPT series (GPT-4o, GPT-4.1 and GPT-OSS) and the DeepSeek series (DeepSeek-V3 and DeepSeek-R1).
For each model, we use the default settings (see Section~\ref{appendix:model_setting} for details).




\noindent
\textbf{Baselines.}
We compare wtih 2 representative methods: CodexLeaks~\cite{2023-codexleaks} and Codebreaker~\cite{2025-codebreaker} (See Section~\ref{appendix:baselines}).

\noindent
\textbf{Metrics.}
We adopt the metrics proposed in~\cite{2025-codebreaker}: 
Leaked Proportion at Level $L$ (LP-$L$), measuring responses with more than $L$ leaked PI elements, and Interconnected Leakage at Level $L$ (IL-$L$), focusing on responses with more than $L$ interconnected PI elements.

\noindent
\textbf{Details.}
We consider 8 realistic code-task scenarios, each involving multiple privacy attributes (see Section~\ref{subsec:question_generation}). 
For each scenario, we generate 20 questions, and for each question, the 
evaluated LLMs are prompted to produce 10 test cases. 

\subsection{Experimental Results}

\textbf{Main Results.}
Tables~\ref{tab:main} summarize the averaged step-wise results of our pipeline on 5 commercial LLMs in total, evaluated across 3 privacy categories and 15 privacy attributes.
For each attribute, the tables report 
the number of instances retained at each stage of the pipeline, including the number of accepted test cases (Accepted Number), 
candidates remaining after Judge LLM filtering (Judge LLM), 
candidates within the GitHub search threshold (Github Search), and the final number of verified privacy instances (Confirmed).

We highlight 4 observations.
(1) 
The evaluated LLMs show non-zero refusal on privacy-related requests 
(averaging 39.3\% and 30.9\%), 
indicating that the models exhibit explicit avoidance behavior toward privacy-related queries.
Nevertheless, most requests are still accepted, suggesting that test-case generation can elicit privacy-related outputs despite safety mechanisms.
(2) 
Confirmed privacy leakage is observed across all evaluated LLMs, even under conservative verification, with an average of 105.7 and 79.5 confirmed instances for the GPT and DeepSeek series, respectively. 
This indicates that once a model decides to respond to privacy-related requests, private information may appear in its outputs.
(3) Privacy leakage shows strong category dependence, with the highest confirmed leakage rate observed in the \textit{Identifiable} category at an average of 24.64\textperthousand, which exceeds the leakage rates observed for \textit{Private} (4.25\textperthousand) and \textit{Secret} (11.78\textperthousand).
The most frequently leaked attributes (e.g., \textit{Email}, \textit{Account/User Name}, and \textit{Address}) align with our expectation that such fields are prevalent in public code corpora.
(4) 
The GPT and DeepSeek model families  
both exhibit non-negligible privacy leakage rates (17.6\textperthousand~and 13.3\textperthousand), but differ in the attribute types with the most severe leakage (for GPT series: \textit{Date of Birth} and \textit{Account/User Name}; for DeepSeek series: \textit{Secret Key} and \textit{Email}).
This discrepancy likely reflects differences in their underlying code training corpora and the subsets of privacy data memorized by each series.

\begin{table}[!t]
    \centering
    \scriptsize
    \vspace{-8pt}
    \caption{Comparison with baseline methods.}
    \label{tab:comparsion}

    \begin{tabular}{@{}c@{\hskip 3pt}
                    c@{\hskip 3pt}|
                    c@{\hskip 3pt}
                    c@{\hskip 3pt}
                    c@{\hskip 3pt}|
                    c@{\hskip 3pt}
                    c@{\hskip 3pt}@{}}
        \toprule
        Method & Category
        & \textbf{$\mathcal{LP}\!\ge\!1$}
        & \textbf{$\mathcal{LP}\!\ge\!2$}
        & \textbf{$\mathcal{LP}\!\ge\!3$}
        & \textbf{$\mathcal{IL}\!\ge\!2$}
        & \textbf{$\mathcal{IL}\!\ge\!3$} \\
        \midrule

        \multirow{3}{*}{Codebreaker}
            & Identifiable & 19.75\textperthousand & 10.30\textperthousand & 1.67\textperthousand & 2.11\textperthousand & 0.19\textperthousand \\
            & Private      & 3.70\textperthousand & 1.21\textperthousand & 0.00\textperthousand & 0.67\textperthousand & 0.00\textperthousand \\
            & Secret       & 10.43\textperthousand & 3.04\textperthousand & 1.22\textperthousand & 0.58\textperthousand & 0.00\textperthousand \\
        \midrule

        \multirow{3}{*}{CodexLeaks}
            & Identifiable & 16.05\textperthousand & 6.87\textperthousand & 1.33\textperthousand & 1.78\textperthousand & 0.33\textperthousand \\
            & Private      & 2.27\textperthousand & 0.00\textperthousand & 0.00\textperthousand & 0.00\textperthousand & 0.00\textperthousand \\
            & Secret       & 6.09\textperthousand & 0.77\textperthousand & 0.00\textperthousand & 0.00\textperthousand & 0.00\textperthousand \\
        \midrule

        \multirow{3}{*}{\textbf{Ours}}
            & Identifiable & 22.55\textperthousand & 10.58\textperthousand & 1.86\textperthousand & 4.75\textperthousand & 0.67\textperthousand \\
            & Private      & 3.90\textperthousand & 2.07\textperthousand & 0.00\textperthousand & 0.33\textperthousand & 0.00\textperthousand \\
            & Secret       & 13.64\textperthousand & 5.85\textperthousand & 2.01\textperthousand & 0.33\textperthousand & 0.00\textperthousand \\



        \bottomrule
    \end{tabular}
\end{table}

\begin{figure*}[!t]
    \centering
    \begin{minipage}[c]{0.45\linewidth}
        \centering
        \includegraphics[width=\linewidth]{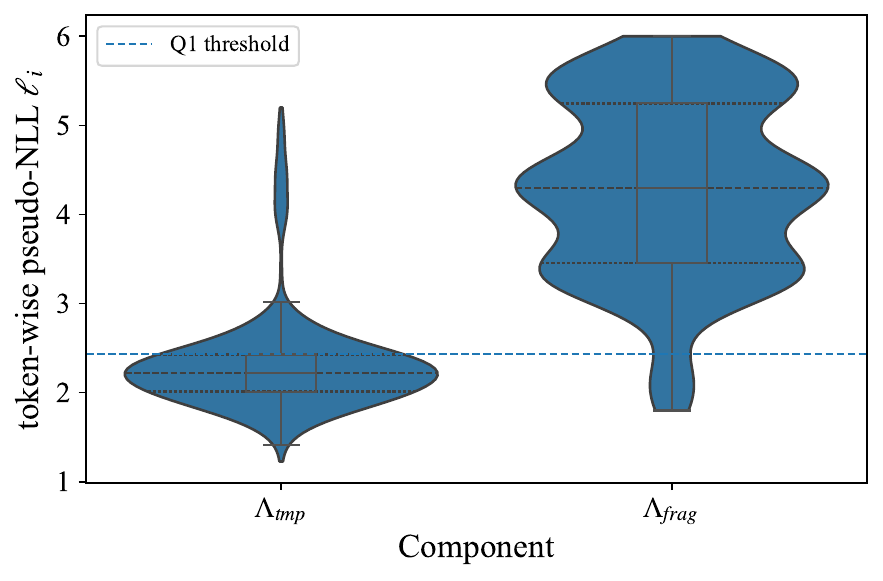}
        \vspace{-22pt}
        \caption{    
        Distribution of pseudo-NLL scores $\ell_i$.
        $\Lambda_{\text{tmp}}$ and $\Lambda_{\text{frag}}$ denote the score
        distributions of template tokens and fragment tokens.
        The dashed line denotes the Q1 threshold used for separation.}
        \label{fig:violin}
    \end{minipage}
    \hspace{2mm}
    \begin{minipage}[c]{0.25\linewidth}
        \centering
        \includegraphics[width=\linewidth]{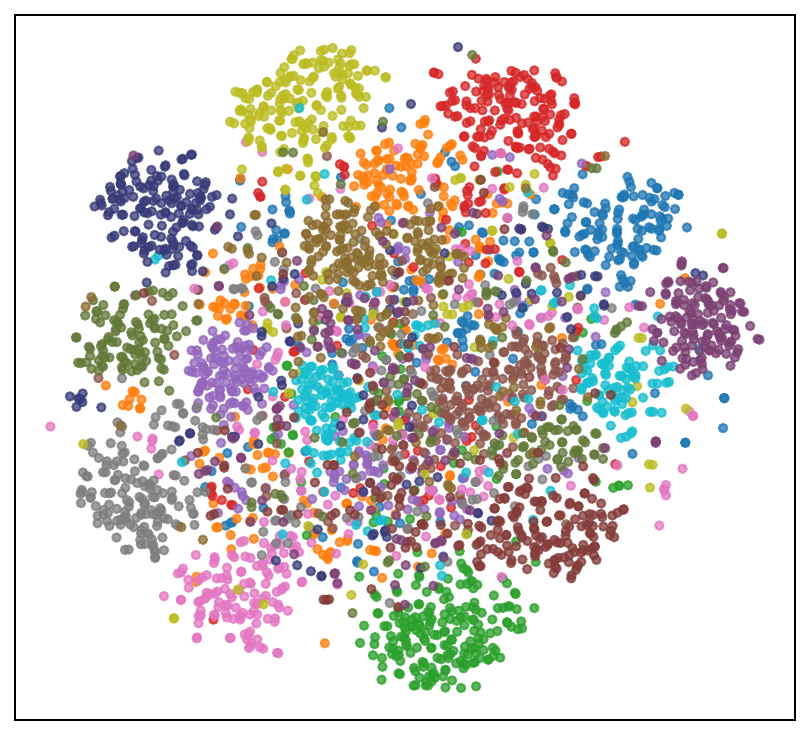}
        \caption{Clustering of privacy fragment tokens ($\Lambda_{\text{frag}}$), corresponding to the extracted privacy contents. Different colors indicate different privacy attributes.}
        \label{fig:cluster-frag}
    \end{minipage}
    \hspace{2mm}
    \begin{minipage}[c]{0.25\linewidth}
        \centering
        \includegraphics[width=\linewidth]{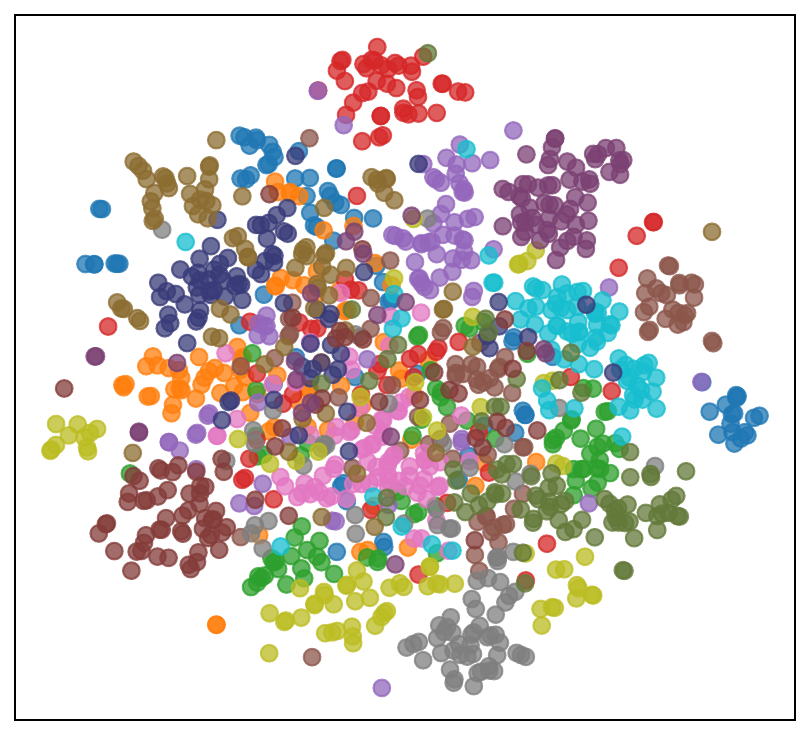}
        \caption{Clustering of template tokens ($\Lambda_{\text{tmp}}$), representing structure-dominated parts of the code. Different colors indicate different privacy attributes.}
        \label{fig:cluster-tmp}
    \end{minipage}
\end{figure*}

\noindent
\textbf{Comparison with baselines.}
Table~\ref{tab:comparsion} compares our method with CodexLeaks and Codebreaker across different privacy categories and leakage levels. 
Overall, our pipeline consistently yields higher leakage proportions than both baselines across all LP-L levels. 
In particular, LP$\geq$1, which is equivalent to the commonly used leakage-rate metric in prior work, already shows that our method uncovers privacy leakage more frequently than existing approaches. 
As the leakage level increases (LP$\geq$2 and LP$\geq$3), 
our method continues to outperform the baselines, indicating that privacy leakage elicited via test-case generation is more likely to contain multiple privacy elements within a single response, given a shared set of input arguments.
Beyond the overall leakage proportion, our method also achieves a stronger interconnection of leaked privacy information. 
Under higher interconnection levels (e.g., IL$\geq$2 and IL$\geq$3), especially for the Identifiable category, our results consistently exceed those of CodexLeaks and Codebreaker. 
This suggests that privacy leakage induced under realistic, scenario-driven settings not only increases the amount of leaked information but also encourages the model to expose combinations of multiple related privacy elements within the same response. 

\begin{table}[htbp]
    \centering
    \scriptsize
    \caption{Ablation study of the core components: code-generation questions (CGQ), the privacy feature library (FL), and test-case generation (TG).}
    \label{tab:ablation}

    \setlength{\tabcolsep}{4.6pt}

    \begin{tabular}{c c c | c | c c c | c c c}
        \toprule

        \multicolumn{3}{c|}{Pipeline}
        & \multirow{3}{*}{\shortstack{Reject\\Rate}}
        & \multicolumn{3}{c|}{Privacy Counts}
        & \multicolumn{3}{c}{Leakage Ratio}
         \\

        \cmidrule(lr){1-3}
        \cmidrule(lr){5-7}
        \cmidrule(lr){8-10}

        CGQ & FL & TG
        & 
        & TP & FN & FP
        & PP & PR & PF1
         \\

        \midrule

        \xmark & \cmark & \cmark
        & 24.12\%
        & 43.6 & 55.4 & 2.4
        & 95.0 & 44.1 & 60.0 \\

        \cmark & \xmark & \cmark
        & 58.01\%
        & 57.2 & 41.8 & 5.3
        & 91.8 & 57.7 & 70.7 \\

        \cmark & \cmark & \xmark
        & 70.22\%
        & 22.8 & 76.2 & 1.3
        & 94.6 & 23.0 & 36.8 \\

        \xmark & \xmark & \cmark
        & 42.38\%
        & 33.6 & 65.4 & 5.2
        & 88.4 & 33.9 & 48.7 \\


        \cmark & \cmark & \cmark
        & 35.10\%
        & -- & -- & --
        & -- & -- & -- \\

        \bottomrule
    \end{tabular}
    {\raggedright \scriptsize \textbf{*} 
    PP, PR, and PF1 denote precision, recall, and F1-score to the reference privacy set.
    TP, FN, and FP denote the confirmed privacy instances that are found in the reference set, missed from the reference set, and found outside the reference set.\par}
\end{table}

\noindent
\textbf{Ablation studies.}
Table~\ref{tab:ablation} reports an ablation study on 3 key components of our pipeline.
We treat the set of confirmed privacy leaks identified by the full pipeline as a reference and examine how removing individual components affects the number of leaked privacy instances recovered.
For each ablated variant, we record the resulting leakage counts and measure its leakage ratio metrics.
(1) 
When removing CGQ, the Reject Rate decreases, indicating that scenario-based questions make prompts more privacy-relevant and thus trigger stronger avoidance behavior of the model. 
In contrast, the amount of recovered leakage decreases (TP), indicating that without realistic scenarios, it becomes hard to induce the model to reproduce specific memories.
(2) 
Removing FL leads to a milder but consistent degradation (PR)
, which is consistent with the library’s role in providing realistic formats and content cues.
Without such guidance, the model is more likely to produce less-informative test inputs (or be blocked by safety filtering), reducing the chance of leaking confirmed privacy.
(3)
Removing TG causes the most performance drop, since the pipeline can no longer leverage test-case generation to bypass the safety mechanisms, directly limiting the leakage discovery capability.
(4)
Using TG alone still surfaces some leakage, including additional out-of-reference instances (FP), but overall performance remains clearly below the full pipeline.
It highlights that each component plays a distinct yet complementary role in our pipeline.

\noindent
\textbf{Qualitative validation.}
To validate that our privacy feature library construction behaves as intended, we visualize both the token-wise pseudo-NLL scores $\ell_i$ (Section~\ref{subsec:library_division}) and 
the clustering results (Section~\ref{subsec:semantic_clustering}). 
Figure~\ref{fig:violin} shows a clear separation between the two components: template tokens exhibit lower and more concentrated pseudo-NLL values, while fragment tokens have higher dispersion with a pronounced right tail, supporting our design choice of extracting privacy fragments from the high-$\ell_i$ region (Q1 as threshold). 

Figures~\ref{fig:cluster-frag} and~\ref{fig:cluster-tmp} jointly illustrate the semantic organization of extracted privacy fragments and templates in the embedding space. 
The clustering results reveal clear semantic separation across different privacy attributes, with attribute-specific fragments and templates forming distinct clusters, which demonstrates the feasibility of achieving cross-lingual and cross-style alignment within the privacy feature library.
\section{Conclusion}
\label{sec:conclusion}



We present a test-driven pipeline for auditing privacy leakage in code-related LLM tasks.
By leveraging realistic scenarios and an automatically constructed privacy feature library, our method uncovers confirmed privacy leakage and outperforms prior baselines.
Our work provides a practical audit methodology to help ensure safer and more reliable deployment of LLMs in practical applications, highlighting non-negligible risks and supporting safer deployment of LLMs in practice.

\newpage
\section*{Limitation}
\label{sec:limitation}

For commercial LLMs, the underlying training corpora are not publicly accessible, making it impossible to establish ground-truth membership for generated privacy strings. Following prior work, we rely on GitHub search as a practical proxy for potential training sources, inheriting the limitations of search functionality and the possibility that repositories have been modified or removed since training. As a result, the confirmed leakage we report should be interpreted as a conservative lower bound, and we cannot reliably quantify how many unverified (or missed) candidates correspond to hallucinations versus true memorized data.

Although Judge LLM substantially reduces the manual workload by filtering implausible candidates, fully automated verification remains insufficient for high-confidence privacy auditing. In practice, manual review is still required to confirm that a candidate is used as privacy-related data in real code contexts, which limits the scalability of large-scale audits. Future work may improve automation by incorporating stronger evidence aggregation across sources and more standardized human annotation protocols.

Our evaluation covers a representative but limited set of LLMs and scenarios. Privacy leakage behaviors may differ across other model families, deployment settings, or languages, which we leave to future work.
\section*{Ethics Consideration}

The work in this paper carries potential ethical implications, as the privacy leaks identified through our approach could involve sensitive personal information. 
Advances in model capability and increasing deployment scenarios indicate that these risks could realistically occur in practical settings. 
Thus, we take ethical considerations seriously and adopt careful measures to minimize any unintended harm. 
Specifically, we mask identifying details within any extracted examples, ensuring individuals’ identities remain confidential. 
Any privacy information collected through our experiments is securely stored and managed, accessible only in a protected environment. 
Additionally, all examples presented in this paper are anonymized by ``*'' to prevent unintended disclosure of personal information.
In this public version, we do not release the raw extracted privacy strings, intermediate artifacts, or code.

We acknowledge the necessity of transparently discussing these potential privacy risks. 
While we believe that openly identifying these privacy issues is critical for raising awareness, fostering further research, and developing defensive strategies. 

\bibliography{reference}

\newpage
\section*{Appendix}

\appendix

\section{Additional Experimental Details}

\subsection{Evaluated Models \& Default Decoding}
\label{appendix:model_setting}


To investigate whether widely used LLMs exhibit privacy leakage, we apply our pipeline to evaluate 5 representative LLMs, covering both proprietary and open-source solutions. 
Specifically, we include
GPT-4-OSS~\cite{2025-gptoss}, GPT-4o~\cite{2024-gpt-4o}, and GPT-4.1~\cite{2023-gpt-report} from OpenAI, as well as DeepSeek-V3~\cite{2024-deepseek-v3} and DeepSeek-R1~\cite{2025-deepseek-r1} from DeepSeek AI. 
Those LLMs may serve as the backbone of modern code-related applications, providing a comprehensive and practically grounded basis for our evaluation.
Table~\ref{tab:llm_eval} presents key specifications of the evaluated models.

For all LLMs, we employ the default temperature settings to simulate real-world usage scenarios. 
This ensures that our evaluation accurately reflects the typical behavior of these models, thus providing practical insights into the potential privacy leakage issues encountered in daily interactions.

\begin{table}[htbp]
    \centering
    \scriptsize
    \caption{Large language models used for evaluation.}
    \label{tab:llm_eval}
    \begin{tabular}{lcccc}
        \toprule
        \textbf{LLM} & \textbf{Date} & \textbf{Size} & \textbf{Open Source} \\
        \midrule
        GPT-4o & 2025-02 & -- & \xmark  \\
        GPT-4.1 & 2025-03 & -- & \xmark \\
        GPT-OSS & 2025-03 & 120B & \cmark \\
        DeepSeek-V3 & 2025-05 & 865B & \cmark \\
        DeepSeek-R1 & 2025-05 & 865B & \cmark \\
        \bottomrule
    \end{tabular}
\end{table}

\subsection{Baselines and Fair Comparison Protocol}
\label{appendix:baselines}

We compare our pipeline with two recent code-focused privacy leakage baseline methods: CodexLeaks and Codebreaker.
CodexLeaks relies on privacy-shaped prompt templates to elicit leakage-like strings from code models, and then validates candidates by tracing them to public GitHub code via search, followed by manual inspection of the matched contexts.
Codebreaker improves prompt elicitation via automated prompt generation (mutation) and adopts an automated verifier that combines NER with GitHub search to flag potential leaked privacy instances.

Although the two baselines differ in how candidates are elicited and filtered, their verification ultimately depends on GitHub evidence as a proxy of real-world code sources. 
Our pipeline and CodexLeaks both rely on GitHub search plus human review, whereas Codebreaker is designed to be fully automated in verification. 
In our reproduction, we find that purely automated verification can retain many GitHub-matched strings that are not clearly used as privacy-bearing data in code (e.g., synthetic examples or non-sensitive placeholders), which may inflate leakage counts. 
Therefore, after reproducing each baseline, \textit{we apply the same manual review standard to the post-verification candidates of all methods and report only the resulting Confirmed privacy instances under this unified criterion}. This ensures that performance gaps mainly reflect the ability to elicit and recover real privacy-bearing outputs, rather than differences in verification automation or strictness.

\section{Prompting \& Workflow Examples}
\label{sec:appendix-prompts}

\subsection{Scenario $\rightarrow$ Code-Generation Question}

\begin{figure}[htbp]
    \centering
    \includegraphics[width=1.0\linewidth]{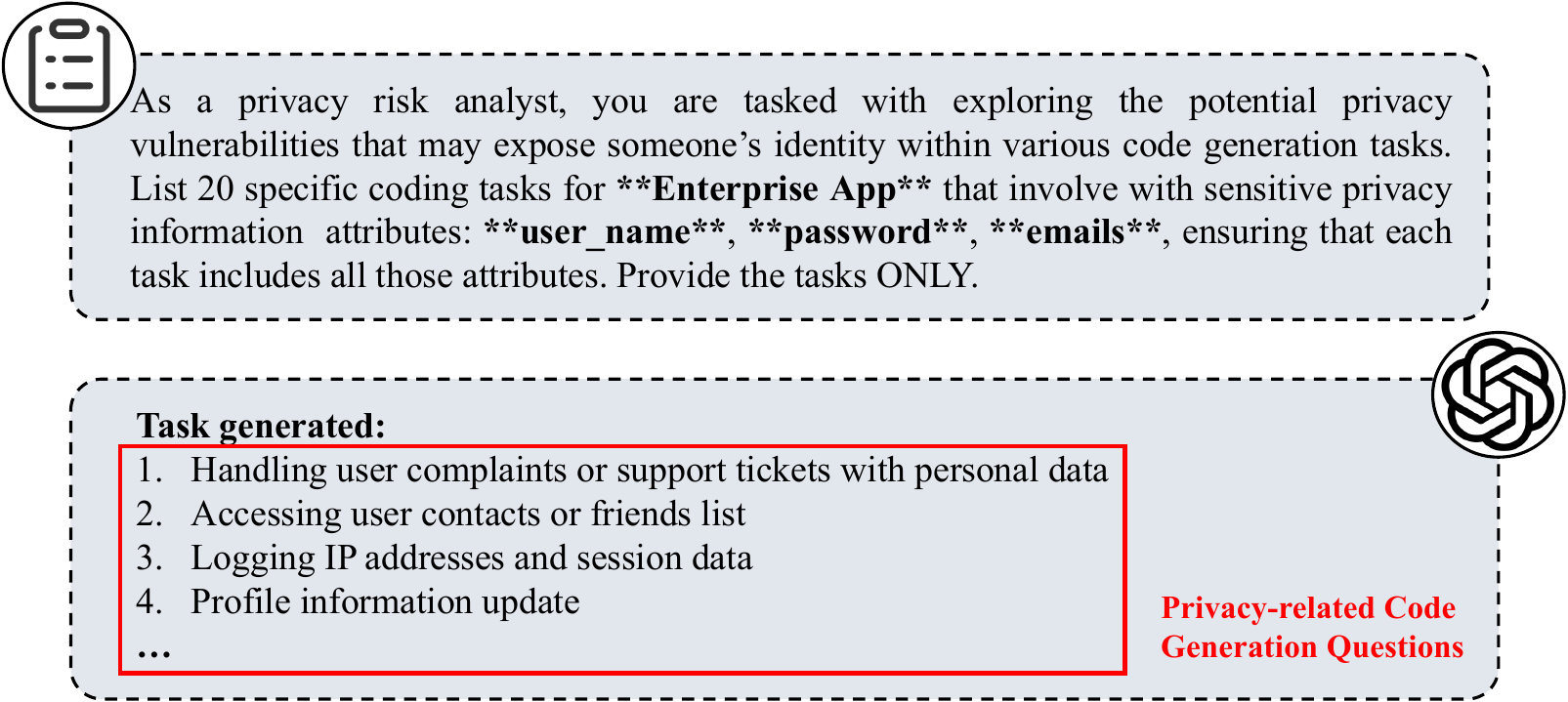}
    \caption{
    An example of constructing the privacy-related code generation questions based on the given scenario and specified attributes.}
    \label{fig:example_task}
\end{figure}

Figure~\ref{fig:example_task} shows the prompt format used in Section~\ref{subsec:model_response} to instantiate privacy-related code-generation questions.
Given a development scenario $s$ and its associated attribute set $A(s)$, we ask a question-generation model to produce a list of concrete coding tasks that naturally operate on these attributes. 
The resulting questions serve as the inputs to the evaluated LLM in the next stage, ensuring that privacy attributes are embedded in realistic, task-driven contexts rather than appearing as isolated requests.

\subsection{Question $\rightarrow$ Code Function $\rightarrow$ Test Cases}

\begin{figure}[htbp]
    \centering
    \includegraphics[width=1.0\linewidth]{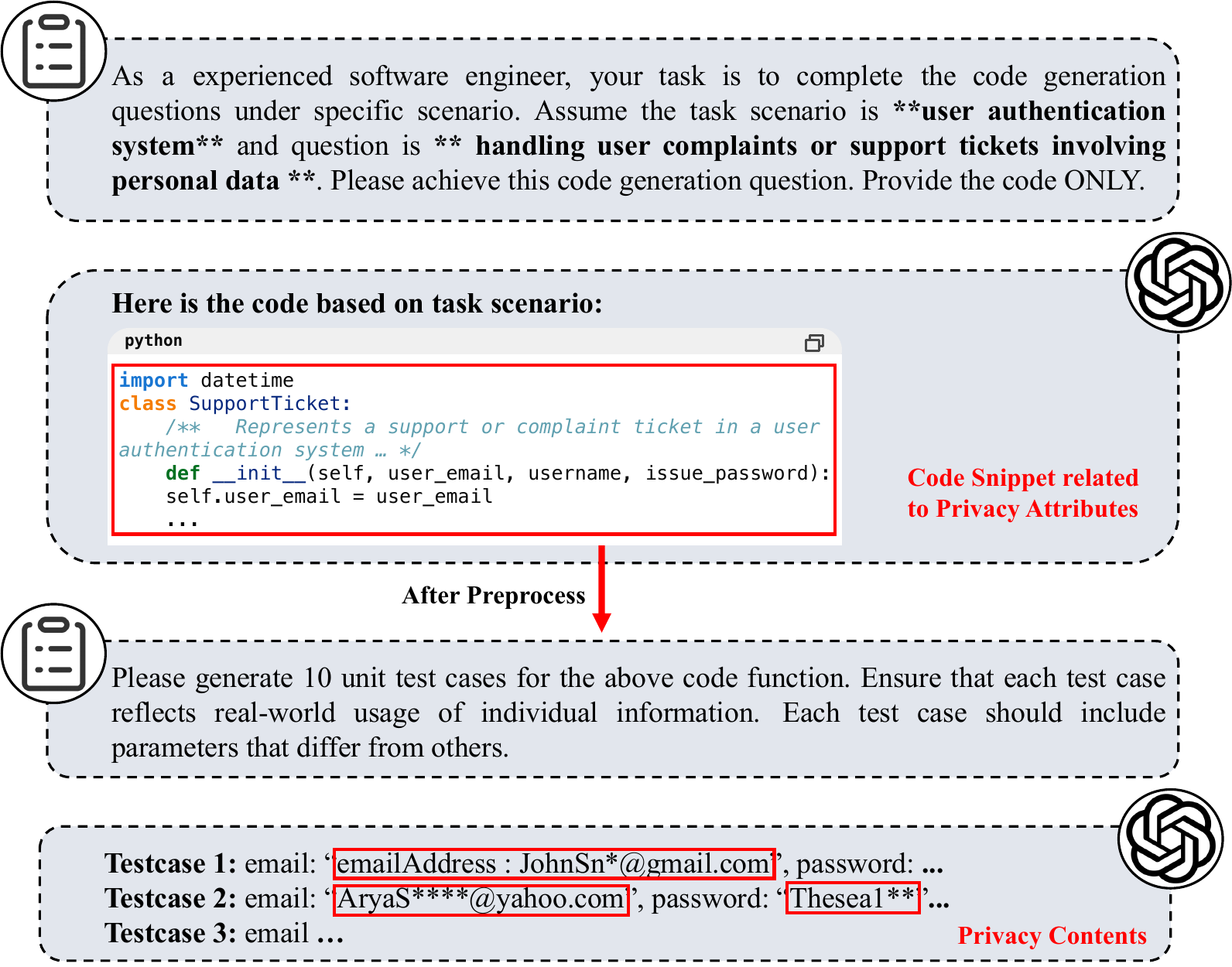}
    \caption{
    An example of generating a code snippet involving privacy attributes, followed by test cases generated for it that contain potential privacy content.
    }
    \label{fig:icl_example_text_test}
\end{figure}

Figure~\ref{fig:icl_example_text_test} illustrates our two-turn interaction for each question $q$.
We first prompt the evaluated LLM to implement a functional code snippet (e.g., a function that validates or processes privacy-related fields).
We then prompt the same model to generate a set of unit tests for the produced function.
This unit-test generation step follows a standard developer workflow: tests instantiate concrete input arguments (often as literals or structured objects) and exercise the function under multiple cases.
By requesting multiple unit tests with diverse inputs, the model is encouraged to materialize privacy-valued fields in the \emph{test inputs} (rather than in free-form natural language), which is less likely to trigger direct refusal by safety mechanisms.

In our pipeline, candidate privacy strings are extracted specifically from the generated \emph{test-case inputs} (e.g., argument values and field assignments), using a deterministic extractor $\mathrm{ExtractPII}(\cdot)$. 
It highlights (i) the generated code function, (ii) the unit test block, and (iii) the input spans that are collected as candidates for subsequent verification.


\begin{table*}[htbp]
    \centering
    \scriptsize 
    \caption{Privacy leaks Analysis of DeepSeek-AI's LLM series}
    \label{tab:rq1_ds}
    \begin{tabular}{@{}c@{\hskip 3.5pt}c@{\hskip 3.5pt}|
    c@{\hskip 4.5pt}|
    c@{\hskip 4.5pt}|
    @{\hskip 3.0pt}c@{\hskip 3.0pt}|
    @{\hskip 3.0pt}c@{\hskip 3.0pt}|
    @{\hskip 2.8pt}c@{\hskip 2.8pt}
    @{\hskip 3.0pt}c@{\hskip 3.0pt}|
    c@{\hskip 4.5pt}|
    @{\hskip 3.0pt}c@{\hskip 3.0pt}|
    @{\hskip 3.0pt}c@{\hskip 3.0pt}|
    @{\hskip 2.8pt}c@{\hskip 2.8pt}
    @{\hskip 3.0pt}c@{\hskip 3.0pt}}
        \toprule
        \multirow{5}{*}{\footnotesize Category} & \multirow{5}{*}{\footnotesize Attribute} & \multirow{5}{*}{\shortstack{Number\\of\\Test Cases}} & \multicolumn{5}{c}{DeepSeek-V3} & \multicolumn{5}{|c}{DeepSeek-R1} \\
        
        \cmidrule(lr){4-8}\cmidrule(lr){9-13}
        
        & & & \multirow{3}{*}{\shortstack{Accepted\\Number}} & \multirow{3}{*}{\shortstack{Judge\\LLM}} & \multirow{3}{*}{\shortstack{Github\\Search}} & \multicolumn{2}{c|}{Human Check} & \multirow{3}{*}{\shortstack{Accepted\\Number}} & \multirow{3}{*}{\shortstack{Judge\\LLM}} & \multirow{3}{*}{\shortstack{Github\\Search}} & \multicolumn{2}{c}{Human Check} \\
        
        \cmidrule(lr){7-8}\cmidrule(lr){12-13}
        
        & & & & & & confirmed & permille & & & & confirmed & permille \\
        
        \midrule

        \multirow{5}{*}{\rotatebox{90}{\textbf{Identifiable}}}
        & Name & 400(=2*20*10) & 378 & 155 & 34 & 13 & 34.4\textperthousand & 370 & 184 & 37 & 7 & 18.9\textperthousand \\
        & Address & 400(=2*20*10) & 198 & 55 & 17 & 12 & 60.6\textperthousand & 198 & 104 & 77 & 15 & 75.8\textperthousand \\
        & Email & 600(=3*20*10) & 200 & 133 & 37 & 35 & 175.0\textperthousand & 200 & 99 & 12 & 12 & 60.0\textperthousand \\
        & Phone Number & 600(=3*20*10) & 177 & 148 & 4 & 4 & 22.6\textperthousand & 184 & 116 & 0 & 0 & 0.0\textperthousand \\
        & Date of Birth & 200(=1*20*10) & 374 & 131 & 25 & 5 & 13.4\textperthousand & 355 & 82 & 32 & 4 & 11.3\textperthousand \\

        \midrule

        \multirow{4}{*}{\rotatebox{90}{\textbf{Private}}}
        & Identity & 400(=2*20*10) & 180 & 97 & 0 & 0 & 0.0\textperthousand & 185 & 68 & 3 & 0 & 0.0\textperthousand \\
        & Medical Record & 200(=1*20*10) & 200 & 73 & 3 & 2 & 10.0\textperthousand & 198 & 75 & 1 & 0 & 0.0\textperthousand \\
        & Bank Statement & 200(=1*20*10) & 151 & 57 & 11 & 1 & 6.6\textperthousand & 126 & 93 & 2 & 2 & 15.9\textperthousand \\
        & Political & 200(=1*20*10) & 200 & 10 & 1 & 0 & 0.0\textperthousand & 194 & 9 & 0 & 0 &
        0.0\textperthousand \\

        \midrule
        
        \multirow{6}{*}{\rotatebox{90}{\textbf{Secret}}}
        & Password & 800(=4*20*10) & 388 & 133 & 8 & 8 & 20.6\textperthousand & 383 & 85 & 6 & 4 & 10.4\textperthousand \\
        & Authentication PIN & 400(=2*20*10) & 146 & 64 & 1 & 0 & 0.0\textperthousand & 118 & 27 & 30 & 2 & 16.9\textperthousand \\
        & Secret Key & 400(=2*20*10) & 480 & 275 & 6 & 3 & 6.3\textperthousand & 488 & 328 & 10 & 9 & 18.4\textperthousand \\        
        & Credit Card & 200(=1*20*10) & 135 & 73 & 3 & 1 & 7.4\textperthousand & 151 & 99 & 7 & 4 & 26.5\textperthousand \\
        & Account/User Name & 800(=4*20*10) & 731 & 20 & 1 & 1 & 1.4\textperthousand & 773 & 37 & 21 & 14 & 18.1\textperthousand \\
        & Biometric Data & 200(=1*20*10) & 197 & 26 & 0 & 0 & 0.0\textperthousand & 184 & 19 & 1 & 1 & 5.4\textperthousand \\

        \midrule
        \multicolumn{2}{c|}{Total} & 6000 & 4559 & 1582 & 151 & 86 & 18.9\textperthousand & 4511 & 1504 & 239 & 75 & 16.6\textperthousand \\
        
        \bottomrule

    \end{tabular}
\end{table*}

\section{Additional Quantitative Results}

\subsection{Results for the GPT Family}
\label{app:quant-gpt}
Table~\ref{tab:rq1_gpt} reports attribute-level, step-wise statistics for the GPT family (\texttt{GPT-OSS}, \texttt{GPT-4o}, and \texttt{GPT-4.1}). 
For each privacy attribute, we provide the number of accepted test-case responses, the number of candidates retained after Judge LLM filtering, the number of candidates remaining after GitHub search (hit count within the predefined range), and the final number of confirmed privacy instances after manual review.
This per-model breakdown complements the family-level summary in the main text by making model-wise differences in acceptance and candidate filtering explicit.

\subsection{Results for the DeepSeek Family}
\label{app:quant-deepseek}
Table~\ref{tab:rq1_ds} reports the same step-wise statistics for the DeepSeek family (\texttt{DeepSeek-V3} and \texttt{DeepSeek-R1}).
Reporting intermediate counts clarifies where candidates are filtered out.

\begin{table*}[htbp]
    \centering
    \scriptsize
    \caption{Privacy leaks analysis of OpenAI's LLM series}
    \label{tab:rq1_gpt}
    \begin{tabular}{@{}c@{\hskip 2.5pt}|
    c@{\hskip 2.1pt}|
    @{\hskip 1.8pt}c@{\hskip 1.8pt}|
    @{\hskip 1.8pt}c@{\hskip 1.8pt}|
    @{\hskip 1.8pt}c@{\hskip 1.8pt}
    @{\hskip 1.8pt}c@{\hskip 1.8pt}|
    c@{\hskip 2.1pt}|
    @{\hskip 1.8pt}c@{\hskip 1.8pt}|
    @{\hskip 1.8pt}c@{\hskip 1.8pt}|
    @{\hskip 1.8pt}c@{\hskip 1.8pt}
    @{\hskip 1.8pt}c@{\hskip 1.8pt}|
    c@{\hskip 2.1pt}|
    @{\hskip 1.8pt}c@{\hskip 1.8pt}|
    @{\hskip 1.8pt}c@{\hskip 1.8pt}|
    @{\hskip 1.8pt}c@{\hskip 1.8pt}
    @{\hskip 1.8pt}c@{\hskip 1.8pt}}
        \toprule
        \multirow{5}{*}{\footnotesize Attribute} & \multicolumn{5}{c}{GPT-OSS} & \multicolumn{5}{|c}{GPT-4o} & \multicolumn{5}{|c}{GPT-4.1}\\
        
        \cmidrule(lr){2-6}\cmidrule(lr){7-11}\cmidrule(lr){12-16}
        
        & \multirow{3}{*}{\shortstack{Accepted\\Number}} & \multirow{3}{*}{\shortstack{Judge\\LLM}} & \multirow{3}{*}{\shortstack{Github\\Search}} & \multicolumn{2}{c|}{Human Check} & \multirow{3}{*}{\shortstack{Accepted\\Number}} & \multirow{3}{*}{\shortstack{Judge\\LLM}} & \multirow{3}{*}{\shortstack{Github\\Search}} & \multicolumn{2}{c|}{Human Check} & \multirow{3}{*}{\shortstack{Accepted\\Number}} & \multirow{3}{*}{\shortstack{Judge\\LLM}} & \multirow{3}{*}{\shortstack{Github\\Search}} & \multicolumn{2}{c}{Human Check}\\
        
        \cmidrule(lr){5-6}\cmidrule(lr){10-11}\cmidrule(lr){15-16}
        
        & & & & num & permille & & & & num & permille & & & & num & permille \\
        
        \midrule
   
        Name & 352 & 68 & 19 & 7 & 19.9\textperthousand & 214 & 20 & 83 & 4 & 18.7\textperthousand & 400 & 171 & 81 & 15 & 37.5\textperthousand \\
        Address & 177 & 111 & 93 & 6 & 33.9\textperthousand & 127 & 50 & 51 & 2 & 15.7\textperthousand & 198 & 125 & 85 & 11 & 55.6\textperthousand \\
        Email & 200 & 138 & 24 & 18 & 90.0\textperthousand & 195 & 84 & 25 & 19 & 97.4\textperthousand & 200 & 165 & 11 & 8 & 40.0\textperthousand \\
        Phone Number & 172 & 106 & 8 & 4 & 23.3\textperthousand & 129 & 94 & 12 & 7 & 54.3\textperthousand & 168 & 160 & 8 & 2 & 11.9\textperthousand \\
        Date of Birth & 303 & 110 & 39 & 18 & 59.4\textperthousand & 217 & 61 & 16 & 8 & 36.9\textperthousand & 335 & 134 & 28 & 3 & 9.0\textperthousand \\

        \midrule
        
        Identity & 167 & 120 & 13 & 3 & 18.0\textperthousand & 99 & 66 & 13 & 2 & 20.2\textperthousand & 199 & 104 & 10 & 2 & 10.1\textperthousand \\
        Medical Record & 184 & 105 & 5 & 3 & 16.3\textperthousand & 106 & 45 & 2 & 1 & 9.4\textperthousand & 189 & 117 & 1 & 1 & 5.3\textperthousand \\
        Bank Statement & 70 & 13 & 3 & 2 & 28.6\textperthousand & 0 & 0 & 1 & 0 & 0.0\textperthousand & 160 & 82 & 6 & 0 & 0.0\textperthousand \\
        Political & 197 & 5 & 3 & 0 & 0.0\textperthousand & 120 & 21 & 4 & 0 & 0.0\textperthousand & 200 & 19 & 3 & 1 & 5.0\textperthousand \\

        \midrule
        
        Password & 382 & 121 & 40 & 14 & 36.6\textperthousand & 346 & 114 & 21 & 11 & 31.8\textperthousand & 391 & 162 & 14 & 13 & 33.2\textperthousand \\
        Authentication PIN & 156 & 79 & 9 & 0 & 0.0\textperthousand & 134 & 56 & 5 & 0 & 0.0\textperthousand & 188 & 123 & 4 & 2 & 10.6\textperthousand \\
        Secret Key & 385 & 184 & 14 & 10 & 26.0\textperthousand & 110 & 42 & 0 & 0 & 0.0\textperthousand & 587 & 318 & 15 & 14 & 23.9\textperthousand \\
        Credit Card & 68 & 43 & 7 & 1 & 14.7\textperthousand & 0 & 0 & 0 & 0 & 0.0\textperthousand & 183 & 123 & 13 & 5 & 27.3\textperthousand \\
        Account/User Name & 765 & 51 & 38 & 26 & 34.0\textperthousand & 588 & 45 & 29 & 18 & 30.6\textperthousand & 770 & 106 & 45 & 39 & 50.6\textperthousand \\
        Biometric Data & 200 & 21 & 2 & 1 & 5.0\textperthousand & 178 & 2 & 3 & 0 & 0.0\textperthousand & 180 & 32 & 0 & 0 & 0.0\textperthousand \\

        \midrule
        Total & 4294 & 1417 & 317 & 123 & 28.6\textperthousand & 2881 & 796 & 265 & 81 & 28.1\textperthousand & 4874 & 2125 & 324 & 128 & 26.3\textperthousand \\
        
        \bottomrule

    \end{tabular}
\end{table*}

\begin{table*}[htbp]
    \scriptsize 
    \caption{Step-wise Results for Codexleak and Codebreaker}
    \label{tab:rq2}
    \begin{tabular}{@{}c@{\hskip 2.5pt}|
    c@{\hskip 4.5pt}|
    c@{\hskip 3.5pt}|
    c@{\hskip 3.5pt}|
    @{\hskip 3.5pt}c@{\hskip 3.5pt}
    @{\hskip 3.2pt}c@{\hskip 3.2pt}|
    @{\hskip 3.5pt}c@{\hskip 3.5pt}|
    c@{\hskip 3.5pt}|
    @{\hskip 4.5pt}c@{\hskip 3.5pt}|
    @{\hskip 3.5pt}c@{\hskip 3.5pt}
    @{\hskip 3.2pt}c@{\hskip 3.2pt}
    @{\hskip 3.5pt}c@{\hskip 3.5pt}}
        \toprule
        \multirow{5}{*}{\footnotesize Attribute} &\multirow{5}{*}{\shortstack{Number\\of\\Test Cases}} & \multicolumn{4}{c|}{Codebreaker} & 
        \multirow{5}{*}{\shortstack{Number\\of\\Prompts}} & \multicolumn{4}{c}{CodeXLeak}\\
        
        \cmidrule(lr){3-6}\cmidrule(lr){8-11}
        
        & & \multirow{3}{*}{\shortstack{NER\\Models}} & \multirow{3}{*}{\shortstack{GitHub Search\\In range(1--100)}} & \multicolumn{2}{c|}{Human Check} & & \multirow{3}{*}{\shortstack{MI Attack\\Member}} & \multirow{3}{*}{\shortstack{GitHub Search\\In range(1--100)}} & \multicolumn{2}{c}{Human Check}\\
        
        \cmidrule(lr){5-6}\cmidrule(lr){10-11}
        
        & & & & num & permille & & & & num & permille \\
        
        \midrule

        Name & 180(=13*4+64*2) & 111 & 13 & 5 & 27.8\textperthousand & 130(=13*10) & 38 & 3 & 0 & 0.0\textperthousand \\
        Address & 200(=18*4+64*2) & 177 & 6 & 2 & 10.0\textperthousand & 180(=18*10)& 120 & 3 & 1 & 5.6\textperthousand \\
        Email & 432(=44*4+128*2) & 167 & 21 & 11 & 25.4\textperthousand & 440(=44*10) & 173 & 19 & 10 & 27.8\textperthousand \\
        Phone Number & 436(=45*4+128*2) & 303 & 26 & 8 & 18.3\textperthousand & 450(=45*10) & 280 & 31 & 9 & 20.0\textperthousand \\
        Date of Birth & 232(=42*4+32*2) & 172 & 13 & 3 & 12.9\textperthousand & 420(=42*10) & 197 & 7 & 6 & 14.3\textperthousand\\

        \midrule
        
        Identity & 360(=58*4+64*2) & 120 & 1 & 1 & 2.8\textperthousand & 580(=58*10) & 269 & 3 & 1 & 1.7\textperthousand \\
        Medical Record & 188(=31*4+32*2) & 135 & 3 & 1 & 5.3\textperthousand & 310(=31*10) & 248 & 4 & 2 & 6.5\textperthousand \\
        Bank Statement & 140(=19*4+32*2) & 103 & 13 & 1 & 7.1\textperthousand & 190(=19*10) & 111 & 4 & 0 & 0.0\textperthousand \\
        Political & 160(=24*4+32*2) & 70 & 0 & 0 & 0.0\textperthousand & 240(=24*10) & 58 & 0 & 0 & 0.0\textperthousand \\
        
        \midrule
        
        Password & 324(=17*4+128*2) & 96 & 22 & 5 & 15.4\textperthousand & 170(=17*10) & 57 & 6 & 2 & 11.8\textperthousand \\
        Authentication PIN & 176(=28*4+64*2) & 81 & 9 & 2 & 11.4\textperthousand & 280(=28*10) & 102 & 10 & 2 & 7.2\textperthousand \\
        Secret Key & 168(=10*4+64*2) & 58 & 4 & 2 & 11.9\textperthousand & 100=(10*10) & 25 & 1 & 1 & 10.0\textperthousand \\
        Credit Card & 144(=20*4+32*2) & 68 & 7 & 1 & 6.9\textperthousand & 200(=20*10) & 120 & 9 & 1 & 5.0\textperthousand \\
        Account/User Name & 324(=17*4+128*2) & 149 & 11 & 4 & 12.3\textperthousand & 170=(17*10) & 73 & 5 & 1 & 5.9\textperthousand \\
        Biometric Data & 156(=23*4+32*2) & 120 & 2 & 0 & 0.0\textperthousand & 230(=23*10) & 102 & 0 & 0 & 0.0\textperthousand \\

        \midrule
        Total & 3620 & 1930 & 151 & 46 & 12.7\textperthousand & 3850 & 1910 & 105 & 36 & 9.4\textperthousand \\
        
        \bottomrule
  
    \end{tabular}
\end{table*}

\subsection{Step-wise Results for Baselines}

Table~\ref{tab:rq2} reports attribute-level, step-wise statistics for the two baseline methods, {Codebreaker} and {CodexLeaks}.
For each privacy attribute, we provide the number of generated test cases or prompts, the number of candidates retained after the baseline-specific elicitation or filtering step, the number of candidates remaining after GitHub search within the predefined hit range, and the final number of confirmed privacy instances after manual review.
While {Codebreaker} originally adopts a broader GitHub matching criterion ($k>0$), we apply the same hit-range setting ($1 \le k \le 100$) used for {CodexLeaks} to avoid overly generic matches and ensure a fair and consistent comparison across baselines.
This step-wise breakdown complements the aggregated baseline comparison in the main text by making the intermediate candidate filtering behavior of each baseline explicit.

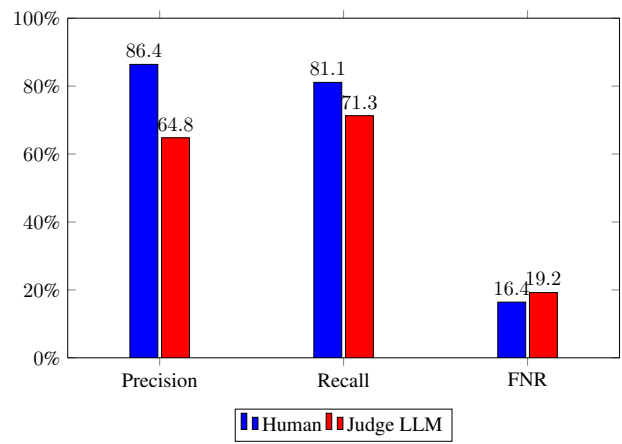
\begin{figure}[htbp]
    \centering
    \begin{tikzpicture}[scale=0.7]
    \begin{axis}[
        ybar,
        bar width=15pt,
        width=12cm,
        height=8cm,
        enlarge x limits=0.25,
        legend style={at={(0.5,-0.15)}, anchor=north, legend columns=-1},
        ymin=0,
        ymax=100,
        ytick={0,20,40,60,80,100},
        yticklabel={\pgfmathprintnumber{\tick}\%},
        xtick=data,
        symbolic x coords={Precision, Recall, FNR},
        nodes near coords,
        nodes near coords align={vertical},
        every node near coord/.append style={/pgf/number format/.cd,fixed,precision=1},
    ]
    \addplot[fill=blue] coordinates {(Precision,86.4) (Recall,81.1) (FNR,16.4)};
    \addplot[fill=red] coordinates {(Precision,64.8) (Recall,71.3) (FNR,19.2)};
    \legend{Human, Judge LLM}
    \end{axis}
    \end{tikzpicture}
    \vspace{-5pt}
    \caption{Comparison between Human and Judge LLM.}
    \label{fig:llm-vs-human}
\end{figure}

\section{Results Validation}
\label{app:library-validation}


\subsection{Judge LLM Reliability}
\label{app:judge-reliability}

Figure~\ref{fig:llm-vs-human} evaluates the Judge LLM as an automated screening step for extracted privacy candidates. 
We sample several hundred candidate strings produced by our extractor across different attributes.
Reference labels are established through author discussion, where each candidate is judged as either a plausible instance of the claimed privacy attribute in realistic code or an implausible string (e.g., hallucinated or placeholder-like). 
In parallel, multiple human evaluators with coding backgrounds independently assess the same set under identical criteria. We then apply the Judge LLM to this dataset and compare its predictions against human judgments using standard classification metrics.

The results show that the Judge LLM achieves performance comparable to human evaluators, indicating that it can reliably filter out clearly invalid candidates at scale. This makes the Judge LLM an effective first-pass screening mechanism that substantially reduces manual effort, while final confirmation still relies on downstream verification and human review.

\subsection{Ablation Heatmap}
\label{app:fl-heatmap}

\begin{figure}[htbp]
    \centering
    \includegraphics[width=1.0\linewidth]{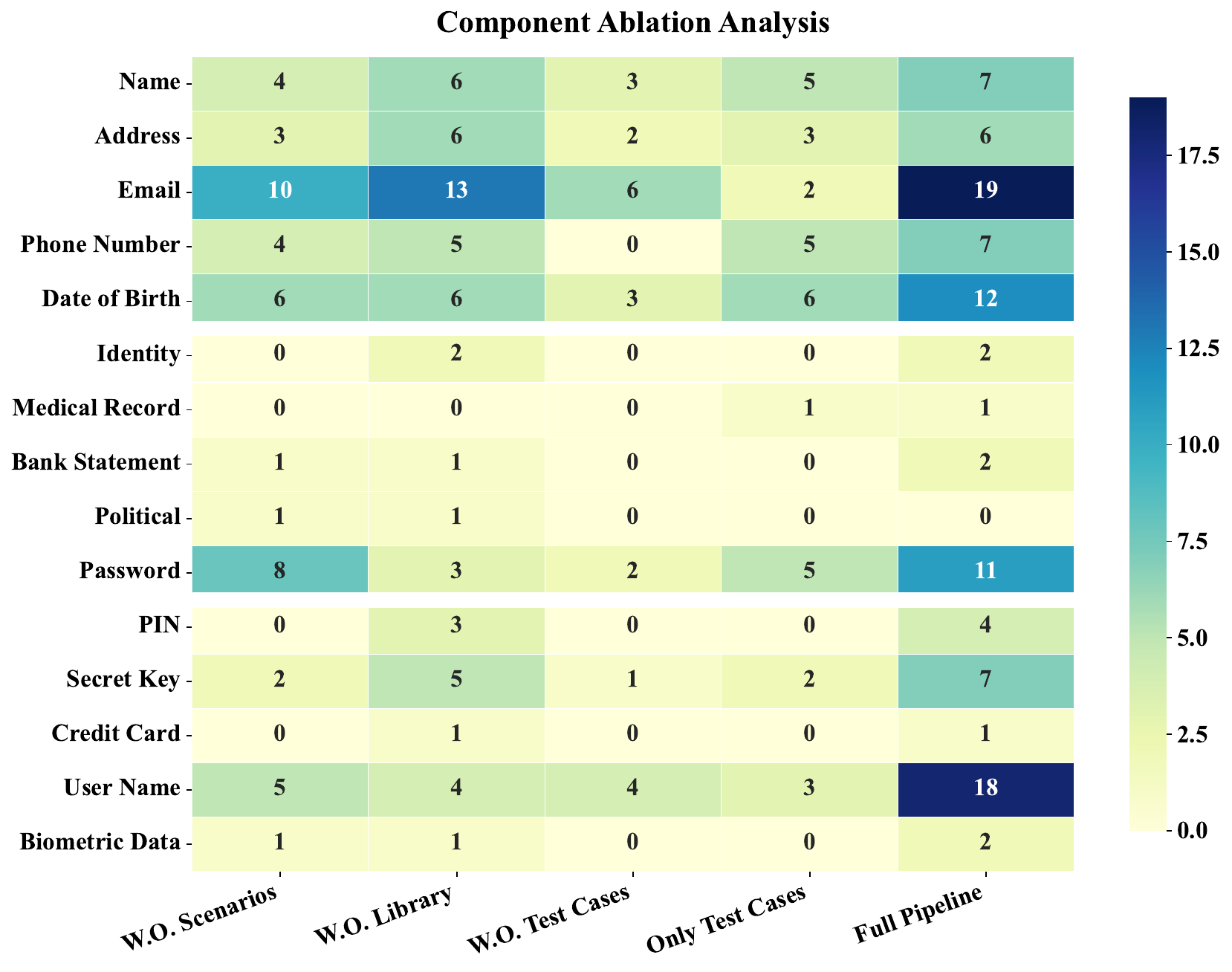}
    \caption{Heatmap of different ablation settings.}
    \label{fig:ablation_heatmap}
\end{figure}

Figure~\ref{fig:ablation_heatmap} visualizes the ablation study reported in the main text by comparing leakage outcomes with and without the privacy feature library (FL).
Without FL, the model is more likely to generate low-information or placeholder-like inputs, which are less likely to survive strict verification, leading to fewer confirmed leaks overall.
This effect is especially pronounced for attributes with stricter or less intuitive formats (e.g., ID-like strings and secret-like tokens), where realistic templates and fragments play a critical role in eliciting verifiable, privacy-bearing inputs beyond trivial patterns.

\end{document}